\documentclass[superscriptaddress,showpacs,amssymb,10pt,reprint,aps,prd,longbibliography,nofootinbib,floatfix]{revtex4-1}

\usepackage{geometry}
\usepackage{graphicx,epsfig,amssymb}
\usepackage{amsmath,amsfonts, times}
\usepackage{bm}

\usepackage{graphicx}
\usepackage{subcaption}
\usepackage{float}

\usepackage{epstopdf}
\usepackage[linktocpage,colorlinks]{hyperref}
\usepackage[usenames]{color}

\usepackage{soul}
\usepackage[utf8x]{inputenc}
\usepackage[usenames]{color}
\usepackage{booktabs}
\usepackage{threeparttable}
\newcommand{\beq}{\begin{equation}}
\newcommand{\eeq}{\end{equation}}
\newcommand{\bqn}{\begin{eqnarray}}
\newcommand{\eqn}{\end{eqnarray}}
\begin{document}
\title{Gravitationally-Induced Photon Entanglement in an FLRW Cosmological Background}
\author{Chi Zhang}
\email{zhangchi3244@gmail.com; Corresponding author}
\affiliation{Department of Physics, Zhejiang Ocean University, Zhejiang, 316022, China}
\begin{abstract}
In order to detect the quantum nature of gravity, the quantum gravity induced entanglement of masses (QGEM) has been proposed both in flat and curved spacetime. In this paper we propose an analogous QGEM protocol using photons produced in astronomical processes as the quantum systems. Unlike massive particles, the gravitational interaction between photons—intrinsically relativistic particles—simultaneously satisfies both the “event” and “system” localities. So it can provide a clear test of whether the gravitational mediator must be nonclassical based on the Local Operations and Classical Communication (LOCC) principle. Although the gravitationally induced entanglement between massless relativistic photons is extremely small, our quantitative calculations clarify the characteristic features of the entanglement induced by the photon's own gravitational field in astronomical process and may inspire other photon entanglement experiments.
\end{abstract}
\date{\today}
\maketitle
\section{Introduction}
The unification of gravity with quantum mechanics remains one of the most profound open questions in theoretical physics \cite{DeWitt:1967uc,Penrose:1996cv,Nandi:2023tfq}. To date a complete theory of quantum gravity is still lacking, and there is no direct experimental evidence of the quantum behavior of the gravitational field. Because the Planck scale is so high, experimental tests have focused on indirect, low-energy schemes. In particular, recent advances in controlling mesoscopic systems have inspired a host of tabletop proposals to witness gravity’s quantum features. Notably, Bose et al., Marletto and Vedral \cite{r12,r13} suggested that two microscale masses—each placed in a spatial superposition—could become entangled via their mutual Newtonian gravitational interaction, providing an LOCC-based test of the gravitational mediator. This ingenious scheme exploits the fact that, in quantum information theory, entanglement cannot arise between two systems through purely Local Operations and Classical Communication (LOCC), so observing an induced entanglement could signal a quantum gravitational field mediator. These and similar quantum gravity induced entanglement of masses (QGEM) ideas have attracted great interest as potential tests of quantum gravity \cite{Carney:2018ofe}. Later, Zhang et al. generalize the QGEM framework to curved spacetime to verify more universal quantum gravity \cite{Zhang:2023llk,Zhang:2023bka}. 

However, it has been pointed out that generating entanglement via gravity does not unambiguously prove that the gravitational field itself is a quantum entity.  Mart\'{\i}n-Mart\'{\i}nez and Perche \cite{Martin-Martinez:2022uio} argue that one can reinterpret the QGEM experiment in quantum-controlled classical field theory unless additional locality constraints are imposed. In particular, the issue of locality in gravitational interactions is subtle. Two notions of locality have been defined \cite{Martin-Martinez:2022uio}: event locality (operations occur at definite spacetime events and cannot produce instantaneous effects at causally disconnected points) and system locality (operations on two quantum systems must be separable). The original QGEM proposal \cite{r12,r13} is manifestly event-local but fails system locality, because its interaction Hamiltonian correlates the two masses in a non-separable way.  Consequently, without both locality criteria, gravity-induced entanglement remains ambiguous.

Besides, while the QGEM proposal offers valuable guidance for testing quantum effect of gravitational field, it targets only non-dynamical Newtonian gravity and previous quantum entanglement experiments have all used massive particles. Building on Feynman's thought experiment and the QGEM framework, recent efforts have extended the analysis into the relativistic regime by studying non-dynamical curved spacetimes induced by a mass in quantum superposition, with the aim of uncovering gravity's quantum properties beyond the Newtonian domain \cite{Kaku:2024lgs,r18,Kabel:2022cje,r20}.

On the other hand, there is growing interest in detecting photon entanglement in the universe. For example, one class of studies used entanglement to reveal the origin of CMB fluctuations \cite{r4,r5,r6,r7}. Another study used entangled photons from the decay of other parent particles in the universe to test quantum mechanics \cite{r8,r9}. There have also been attempts to find experimental ways to unearth entangled photons taking different paths to Earth in astronomical observations \cite{r3}. In the paper \cite{r15}, the variation of quantum field mode functions caused by cosmic expansion gives rise to quantum entangled states in the early universe. The primordial entanglement from the inflaton field in the early period may have been transferred to other fields accessible today, such as the electromagnetic field, making it possible to observe entanglement among individual photons. 

In this paper we investigate whether two photons—prepared in suitable quantum superpositions and propagating in an FLRW background—can become entangled via their mutual gravitational interaction beyond the Newtonian regime, and whether such an effect can serve as a LOCC-based probe of the gravitational mediator. Photons offer several practical advantages for this purpose: they are abundant in the universe, from luminous galaxies to the cosmic microwave background (CMB); for low-energy photons at extremely short distances, the gravitational potential energy can already exceed the higher-order QED interaction energy; they can be detected with high precision; when traveling through low-density regions they can maintain coherence over very large distances.

We show that, under our emission and propagation protocol, both event and system locality can be satisfied. Following the logic of Nandi et al. \cite{r25}, if both locality constraints are enforced and entanglement is observed, the mediator must be non-classical. Our photon-based protocol therefore offers a potentially robust LOCC-style test of gravitational quantization by respecting all relevant locality conditions. 
 
In this paper, we adopt a semi-classical computational framework and assume that general relativity describes spacetime at the photon scale. If this pair of photons in space superposition can indeed become entangled through gravitational mediation within the LOCC framework in the FLRW universe, it will not only broaden experimental targets for us to verify the effect of quantum gravity, but also provide more reference for other photon entanglement experiments in astrophysical settings.

This paper is organized as follows: In section \uppercase\expandafter{\romannumeral2}, we will illustrate in general how photon pairs become entangled under their own gravitational field in FLRW. In section \uppercase\expandafter{\romannumeral3}, we evaluate the decoherence and conversion rate of the spatially superposed photon induced by photon–graviton coupling. In section \uppercase\expandafter{\romannumeral4}, we investigate the variation of the entanglement phase under different initial conditions, and perform a Monte Carlo simulation to examine its statistical properties within the cosmic microwave background. In section \uppercase\expandafter{\romannumeral5}, we will present a conclusion on photon entanglement within the framework of the QGEM mechanism.

In this paper, we adopt the geometric units system, c=G=1, to simplify the formulas, except where physical quantities with units are given in the international system of units (SI).

\section{The QGEM mechanism of photon entanglement in the FLRW universe}

According to current astronomical observations, the universe we inhabit is very well approximated by a spatially flat, homogeneous, and isotropic FLRW cosmological model. The FLRW cosmological  metric in the conformal coordinates with signature $\left( { + , - , - , - } \right)$ is \cite{r10}:
\beq\label{1}
d{s^2} = a{\left( \eta  \right)^2}\left( {d{\eta ^2} - d{x^2} - d{y^2} - d{z^2}} \right),
\eeq
in which $\eta$ is conformal time and \(a\left(  \eta \right)\) is the scale factor. The proper time t of a co-moving observer satisfies $dt = a\left( \eta \right)d\eta$. The scale factor $a(t)$ obeys the Friedmann equation \cite{r10}:
\beq\label{22}
a'{\left( t \right)^2} = \frac{8}{3}\pi a{\left( t \right)^2}\rho \left( t \right).
\eeq
The general density in \eqref{22} is given by \cite{r10}
\beq\label{21}
\rho \left( t \right) = {\rho _{{v_0}}}+\frac{{a_0^3{\rho _{{M_0}}}}}{{a{{\left( t \right)}^3}}} + \frac{{a_0^4{\rho _{{R_0}}}}}{{a{{\left( t \right)}^4}}},
\eeq
where
\beq\label{20}
{\rho _{{v_0}}} = \frac{{3H_0^2{\Omega _\Lambda }}}{{8\pi }},{\rho _{{M_0}}} = \frac{{3H_0^2{\Omega _M}}}{{8\pi }},{\rho _{{R_0}}} = \frac{{3H_0^2{\Omega _R}}}{{8\pi }}.
\eeq
Here $\Omega_{\Lambda}$, $\Omega_M$ and $\Omega_R$ are the density parameters for dark energy, matter and radiation respectively. $H_0$ is the Hubble constant.
The total radiation energy density ${\rho _{{{\rm{R}}_0}}}$ can be derived from the photon energy density ${\rho _{{\gamma _0}}}$:
\beq\label{40}
{\rho _{{{\rm{R}}_0}}} = {\rho _{{\gamma _0}}}\left( {1 + \frac{7}{8}{{\left( {\frac{4}{{11}}} \right)}^{4/3}}{N_{{\rm{eff}}}}} \right),
\eeq
in which $N_{\text{eff}}$ is the effective extra relativistic degrees of freedom in the universe.

There are specific astrophysical mechanisms that could naturally produce coherent multi-path propagation of photon in the universe. Anisotropic media in interstellar dust or black hole accretion disks \cite{r41}, QED vacuum polarization in regions of strong magnetic fields \cite{Lai:2022knd}, and magnetized plasma surrounding pulsars \cite{r40} can lead to polarization-dependent dispersion of photons, which may induce spatial quantum superposition states of photons. Astrophysical masers \cite{Elitzur:1992sb} generate highly coherent radiation which can be described quantum mechanically in well-defined spatial modes. Moreover, gravitational lensing \cite{r42} and other large-scale multipath propagation effects can split a single emission. Assuming their initial quantum superposition state can be expressed as:
\beq\label{31}
\left| {{\Psi _i}} \right\rangle  = \frac{1}{{\sqrt 2 }}(\left| L \right\rangle  + \left| R \right\rangle ),
\eeq
where $ \left| L \right\rangle $  and $ \left| R \right\rangle $ respectively denote the left and right spatial branches of the superposition, respectively. Due to the homogeneity and isotropy of cosmology, we could just set the emission point of the photon as the space origin. Then in one of the conformal frames, a photon labeled 1 emitting along the z axis at conformal time ${\eta _E}$ follows the geodesic:
\beq\label{6}
x\left( \eta  \right) = 0,\; \; y\left( \eta  \right) = 0,\;\; z\left( \eta  \right) = \eta  - {\eta _E}.
\eeq

Because photons travel along null geodesics and their rest mass is zero, the Newtonian-limit approximation used for massive particles  \cite{Zhang:2023llk,Zhang:2023bka} is not applicable here. Instead, we compute photon  metric perturbation then find out phase change step by step. Applying the equivalence principle, we introduce an instantaneous co-moving inertial frame at the spacetime point traversed by photon 1; in this frame the Einstein equation with the photon perturbation reads:
\beq\label{15}
{G_{ab}}  =  - 8\pi T_{ab}^{photon}.
\eeq
In a local region of FLRW universe, if the spatial scale $l$ satisfies \cite{r10}:
\beq\label{2}
lH\left( \eta  \right) \sim \sqrt {\left( {\Delta {x^2} + \Delta {y^2} + \Delta {z^2}} \right)} \frac{{a'\left( \eta  \right)}}{{a\left( \eta  \right)}} \ll 1 ,
\eeq
then the local spacetime can be approximated as Minkowski flat spacetime. In other words, within the spatial region of scale \( l \) in this instantaneous inertial reference frame, the Minkowski flat spacetime approximation works very well.

We assume that the photon could be modelled as a coherent state wave wavepacket. In phase space, a coherent state manifests as a Gaussian wave packet, with its field amplitude and phase approximately well-defined, satisfying the minimum uncertainty relation, and closely resembling a classical electromagnetic field \cite{Scully:1997hcl}. The energy-momentum tensor for a photon traveling along z axis could be expressed in the conformal coordinate as \cite{r16}:
\begin{widetext}
\beq\label{10}
{T^{uv}}\left( {{x^u}} \right) = \frac{{{w_{1\gamma} }a\left( {{\eta _E}} \right)}}{{a{{\left( \eta  \right)}^6}}}\delta \left( x \right)\delta \left( y \right)\delta \left( {z - \eta  + {\eta _E}} \right)\left( {\begin{array}{*{20}{c}}
		1&0&0&1\\
		0&0&0&0\\
		0&0&0&0\\
		1&0&0&1
\end{array}} \right),
\eeq
\end{widetext}
where ${{w_{1\gamma }}}$ is the energy measured by a comoving observer at emission time ${{\eta _E}}$.
To first order, Eq. \eqref{15} reduces to
\beq\label{7}
\frac{1}{2}{\partial _c}{\partial ^c}{h_{ab}} =  - 8\pi T_{ab}^{photon}.
\eeq
This is analogous to an inhomogeneous wave equation, whose retarded solution is \cite{r16}:
\beq\label{8}
{h_{uv}} =  - 4\int {\frac{{\left[ {{{T}_{uv}^{photon}}\left( {{x^u}} \right)} \right]}}{r}dx} dydz,
\eeq
in which the coefficient 4 in the numerator reflects the unique properties of the photon gravitational field. So the perturbation metric is:
\beq\label{11}
\begin{split}
{h_{00}} = {h_{33}} &= - {h_{03}} = - {h_{30}} = A\\
& = - \frac{{4{w_{1\gamma}}a\left( {{\eta _E}} \right)}}{{a{{\left( {{\eta _E} + {z_0}} \right)}^7}\left( {\eta  - {\eta _E} - {z_0}} \right)}},
\end{split}
\eeq
in which
\beq\label{12}
{z_0} = \frac{{{{(\eta  - {\eta _E})}^2} - {x^2} - {y^2} - {z^2}}}{{2\left( {\eta  - {\eta _E} - z} \right)}}.
\eeq
The metric perturbation  exists only inside the future light cone of the emission event, $\eta  - {\eta _E} - \sqrt {{x^2} + {y^2} + {z^2}}  \ge 0$, outside of which there is no perturbation.

\begin{figure}\centering
\includegraphics[scale=0.47]{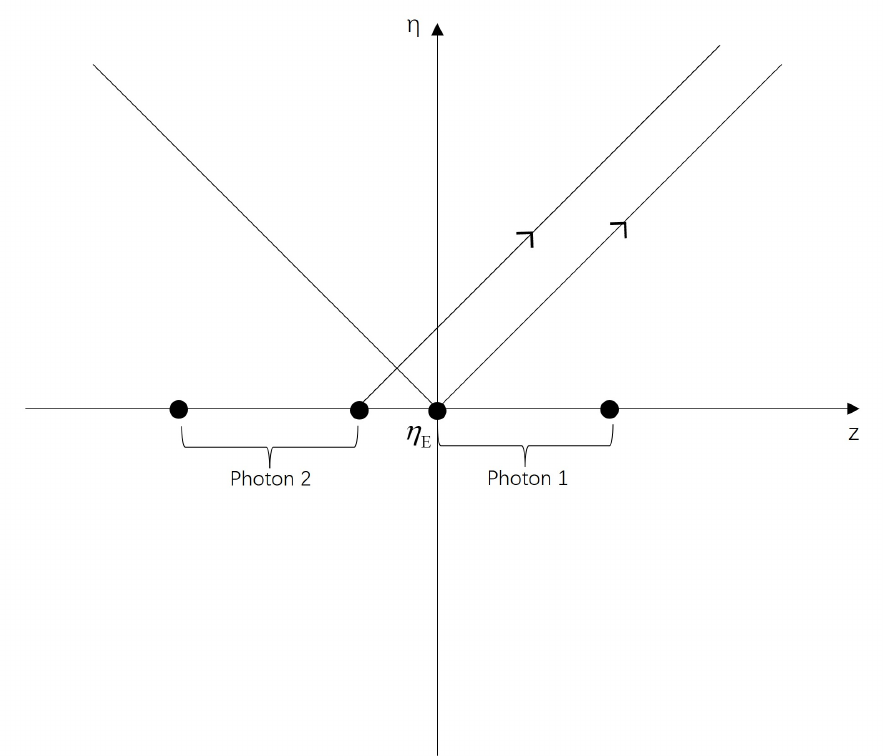}
\caption{Diagram of photon trajectory. All four superposition states move parallel along the Z-axis and we have plotted here only the spacetime trajectory of the state $\left|{RL}\right\rangle_{12}$ we need to consider. In the first order approximation, we just have to think about the unperturbed trajectory of the two photons. In Minkowski spacetime due to the self-stability of the gravitational field of light radiation, the exact trajectory of the second photon inside the light cone of metric perturbation remains a straight line with unit velocity along z-axis.}\label{f1}
\end{figure}

Next, consider a second photon, labeled 2, going off along the z-axis in the vicinity of this spacetime region, such that it is perturbed by the first photon while exerting no influence on the first photon. We assume that photon 2 is also in a spatial superposition state similar to that in Eq. \eqref{31} and the $\left|{RL}\right\rangle_{12}$ are the nearest branches and therefore suffer the largest gravitational effect of the neighboring photon. Their arrangement and motion trajectories have been shown in Fig. \ref{f1}.

In the interacting picture, since only photon 2 is influenced by the gravitational field of photon 1 while photon 1 remains unaffected, the interaction Hamiltonian density contains only the coupling of photon 2 to the gravitational field sourced by photon 1:
\begin{widetext}
\beq\label{33}
{\widehat {\cal H}_I}(x) = {\widehat {\cal H}_{2I}}({x_2}) = - \sum\limits_{{p_2} \in \{ {L_2},{R_2}\} } {\left| {{p_2}} \right\rangle \left\langle {{p_2}} \right| \sqrt { - \det \left( g \right)} T_{{p_2}}^{\mu \nu }({x_2})\hat h_{\mu \nu }^{{{\widehat x}_1}}({x_2})}
\eeq
\end{widetext}
It should be emphasized that Eq. \eqref{33} represents a nontrivial consequence of perturbative quantum gravity. The quantum nature of the graviton leads to a quantum interaction characterized by operator-valued observables \cite{Biswas:2022qto}, in contrast with the classical case. In Refs. \cite{r38, Boccaletti:1969aj}, using an interaction Hamiltonian density analogous to Eq. \eqref{33}, the scattering cross section and the gravitational potential between photons mediated by graviton exchange were derived. Since the resulting cross section is exceedingly small, here we consider only the phase-shift effect at the tree level.

The interaction Hamiltonian is given by:
\beq\label{34}
{{\hat H}_I}(\eta) = \int {d}^3 {x_2} {\widehat {\cal H}_{2I}}({x_2}) \otimes {I_1} = {\widehat H_{2I}}(\eta) \otimes {I_1}
\eeq
Although the interaction Hamiltonian is time-dependent, it commutes with itself at different times according to Eq. \eqref{33}, so the time-ordering operator in the time evolution operator can be omitted. The resulting time-evolution operator is:
\begin{widetext}
\beq\label{35}
{\widehat U_I} = {\widehat U_2} \otimes {\widehat U_1} = \exp \left( {-i\int d \eta{{\widehat H}_{2I}}(\eta)} \right) \otimes {I_1} = \sum\limits_{{p_2} \in \{ {L_2},{R_2}\} } {{e^{i\Delta {\phi _{{p_2}}}\left( {\widehat {{x_1}}} \right)}}\left| {{p_2}} \right\rangle \left\langle {{p_2}} \right|}  \otimes {I_1}
\eeq
\end{widetext}
The factorizability of the evolution operator implies that our system respects system locality. Then according to LOCC, observation of entanglement generated in this setting would then provide strong evidence that the gravitational field carries quantum degrees of freedom \cite{r25}.

Substituting Eq. \eqref{33}, \eqref{34} into Eq. \eqref{35}, the phase change of the photon 2 moving in metric perturbation could  be expressed as:
\beq\label{13}
{\Delta {\phi}} = \frac{1}{{2\hbar }}\int {{h_{uv}}{k^u}d{x^v}},
\eeq
where ${k^u}$ is the unperturbed photon momentum and the integral is integrated along the undisturbed photon geodesic. This is consistent with the phase expression of the photon in a metric perturbation as presented in \cite{r1,r2} through eikonal equation.

From Eq. \eqref{11}, \eqref{12} the $x$- and $y$-directions are equivalent.  Without loss of generality, we choose the transverse separation between photons to lie along the $x$-axis so the transverse separation is $\Delta x$ and denote the longitudinal separation by $\Delta z$. At the locations of the two photons, we establish two instantaneous rest inertial frames of comoving observers. Owing to cosmic expansion, the comoving observer of photon 2 acquires a nonzero relative velocity in the instantaneous inertial frame of photon 1:
\beq\label{4}
{v_x} = \Delta x\frac{{a'\left( \eta  \right)}}{{a\left( \eta  \right)}},\;\;\;\;\;\;{v_z} = \Delta z\frac{{a'\left( \eta  \right)}}{{a\left( \eta  \right)}}
\eeq
The Lorentz transformation matrix between them is therefore:
 \beq\label{5}
\begin{split}
&\Lambda _v^u = \left( {\begin{array}{*{20}{c}}
 		\gamma &{ - {v_x}\gamma }&0&{ - {v_z}\gamma }\\
 		{ - {v_x}\gamma }&{1 + \frac{{v_x^2(\gamma-1)}}{{{\beta ^2}}}}&0&{\frac{{{v_x}{v_z}(\gamma-1)}}{{{\beta ^2}}}}\\
 		0&0&1&0\\
 		{ - {v_z}\gamma }&{\frac{{{v_x}{v_z}(\gamma-1)}}{{{\beta ^2}}}}&0&{1 + \frac{{v_z^2(\gamma-1)}}{{{\beta ^2}}}}
 \end{array}} \right)\\
\\
&\text{where}\;\;\beta = \sqrt {v_x^2 + v_z^2} ,\;\;\;\gamma = \frac{1}{{\sqrt {1-{\beta ^2}} }}.
\end{split}
\eeq
In photon 1’s instantaneous inertial frame the four-momentum and four-displacement of photon 2 transform as:
\beq\label{3}
{k^u}^\prime  = \Lambda _v^u{k^v},\;\;d{x^u}^\prime  = \Lambda _v^ua\left( \eta  \right)d{x^v}
\eeq
When the two photons propagate radially along the z-axis to the Earth observer in a conformally flat spacetime as in Fig.  \ref{f1}, the trajectory of the photon 2 affected by the perturbation is:
\beq\label{26}
x\left( \eta \right) = \Delta x,\;\; y\left( \eta \right) =0,\;\; z\left( \eta  \right) = \eta-{\eta _E} + \Delta z, 
\eeq
Substitute Eq. \eqref{11} and \eqref{4}–\eqref{26} into Eq. \eqref{13}, we obtain:
\begin{widetext}
\beq\label{36}
\begin{split}
\Delta \phi  &= \int\limits_{{\eta _0}}^{{\eta _{now}}} {\frac{{A{w_{2\gamma }}a\left( {{\eta _E}} \right)\left( {2a{(\eta )}^{2}\left( {\sqrt {{1-}\frac{{\left( {\Delta {{x}^{2}}{+}\Delta {{z}^{2}}} \right){a'}{{(\eta )}^{2}}}}{{{a}{{(\eta )}^{2}}}}} {-1}} \right){+}\left( {\Delta {{x}^{2}}{+}\Delta {{z}^{2}}} \right){a'}{{(\eta )}^{2}}} \right)\Delta {x}^{4}}}{{2\hbar {{\left( {\Delta {{x}^{2}}{+}\Delta {{z}^{2}}} \right)}^2}\left( {\left( {\Delta {{x}^{2}}{+}\Delta {{z}^{2}}} \right)a'{{(\eta )}^2} - a{{(\eta )}^2}} \right)}}{d}\eta }\\
&\approx \int\limits_{{\eta _0}}^{{\eta _{now}}} {\frac{{A{w_{2\gamma }}a\left( {{\eta _E}} \right)a'{{\left( \eta  \right)}^4}\Delta {x^4}}}{{8\hbar a{{\left( \eta  \right)}^4}}}d\eta }
\end{split}
\eeq
\end{widetext}
in which ${\eta _0}$ denotes the conformal time of photon 2 at which photon 2 enters the light cone of perturbation. The approximation is made using Eq. \eqref{2} and the leading order term is of fourth order in small quantity.

More generally, the two photons may exhibit a small geodesic deviation. Assuming photon 2 has a small deviation angle $\psi$ in the $x-z$ plane, its geodesic trajectory is given by:
\begin{widetext}
\beq\label{14}
x\left( \eta  \right) = \xi \left( {\eta  - {\eta _E}} \right) + \Delta x,\;\;y\left( \eta  \right) = 0,\;\;z\left( \eta  \right) = \left( {\eta  - {\eta _E}} \right)\sqrt {1 - {\xi ^2}}  + \Delta z
\eeq
\end{widetext}
where $\xi $ is the sine of photon 2’s deviation angle $\psi $ in the $x-z$ plane, i.e. $\xi=\sin \psi$, and is assumed to be small. Analogous to the derivation in Eq. \eqref{36}, the leading order term of both $lH$ and $\xi$ in the entanglement phase between two photons experiencing a small geodesic deviation along the $x$-direction is given by:
\begin{widetext}
\beq\label{28}
\begin{split}
\Delta \phi  \approx &\int\limits_{{\eta _0}}^{{\eta _{now}}} {\frac{{w_{2\gamma} a\left(te\right){A_\xi }}}{{2\hbar }}\left( {\frac{1}{4} {\xi ^4} - \frac{{{\xi ^3} (\Delta x + \xi (\eta-te)) a'\left(\eta \right)}}{{a\left( \eta \right)}}+\frac{{3{\xi ^2} {{(\Delta x + \xi (\eta - te))}^2} a'{{\left( \eta  \right)}^2}}}{{2 a{{\left(\eta \right)}^2}}}} \right.}\\	
&\left. {-\frac{{\xi {{(\Delta x + \xi (\eta-te))}^3} a'{{\left(\eta \right)}^3}}}{{a{{\left( \eta \right)}^3}}} + \frac{{{{(\Delta x + \xi (\eta-te))}^4} a'{{\left(\eta\right)}^4}}}{{4 a{{\left(\eta\right)}^4}}}}\right)d\eta
\end{split}
\eeq
\end{widetext}
	
During propagation the superposed photon pair evolves into a final entangled state that differs only by an overall phase factor:
\beq\label{32}
\left| {{\Psi _f}} \right\rangle  = \frac{1}{2}({\left| {LL} \right\rangle _{12}} + {\left| {LR} \right\rangle _{12}} + {e^{i\Delta \phi }}{\left| {RL} \right\rangle _{12}} + {\left| {RR} \right\rangle _{12}}),
\eeq
in which the subscript 1 and 2 distinguish the two photons. In Eq. \eqref{32}, we neglect the phase changes in the other superposition branches because the  $\left|{RL}\right\rangle_{12}$ branch corresponds to the smallest spatial separation between components and therefore contributes the most significant entanglement phase. So we'll just focus on this branch in the following.

\section{Decoherence induced by photon–graviton coupling}
Our above scheme assumes a priori that the particles are in their spatial superposition states. However, the photon–graviton coupling can mediate photon-graviton conversion in the presence of an external magnetic field \cite{Raffelt:1987im}. Using typical parameters at the epoch of cosmological decoupling \cite{r11}, the oscillation length is: 
\beq\label{50}
{l_{osc}} \approx \frac{{{m_e}^4}}{{{\alpha ^2}{B_e}^2w}} \approx {10^{27}}m
\eeq
which far exceeds the photon propagation distance since decoupling.

Besides, although we neglect other matter fields, gravitational interactions between photons in the CMB can potentially lead to decoherence. The photon in spatial superposition can scatter off blackbody radiation photons through gravitational interactions, the decoherence rate in short-wavelength limit is given by \cite{Boccaletti:1969aj,r43}:
\beq\label{51}
\tau _D^{ - 1} = \int dq \rho \left( q \right) v\left( q \right) {\sigma _{tot}}\left( q \right)
\eeq
where $\rho \left( q \right)$ denotes the photon number density distribution of Planck blackbody radiation, $v\left( q \right)$ is the speed of light and ${\sigma _{tot}}\left( q \right)$ represents the total gravitational scattering cross section between two photons. Since ${\sigma _{tot}}\left( q \right)$ is extremely small \cite{Boccaletti:1969aj}, the decoherence time is ${\tau _D} \sim {10^{97}} s$, which even exceeds the age of universe.

All of these enable us to confidently assume the maintenance of the superposition state throughout the photon’s entire movement.

\section{Influencing factors of photon entanglement and Monte Carlo simulation}
Photon decoupling in the universe occurred at a cosmic age of approximately 379,000 years, corresponding to $z\approx1100$. After decoupling the universe became transparent and decoupled photons could propagate freely. So the earliest photons observable in the universe can be traced back to the epoch of decoupling and the cosmic microwave background radiation photons we observe today were emitted at that epoch. The CMB has a blackbody spectrum whose temperature scales as:
\beq\label{27}
T\left( \eta  \right) = {T_{dec}}\frac{{a\left( {{\eta _{dec}}} \right)}}{{a\left( \eta  \right)}},
\eeq
in which ${T_{dec}}$ is the cosmic temperature at photon decoupling \cite{r10}. According to the black-body radiation law, the  CMB photon phase-space number density is given by:
\beq\label{39}
dn = n\left( {p,\Omega } \right)dpd\Omega  = \frac{{2{p^2}}}{{{h^3}\left( {{e^{p/{k_B}T}} - 1} \right)}}dpd\Omega \eeq
Substituting the present CMB temperature of 2.73 K, the present  number density of CMB photon moving in the same direction is 33 ${cm}^{-3}$. Because the expansion of the universe causes the spatial volume to scale as \((1 + z)^{-3}\), the photon number density scales with redshift as:
\beq\label{37}
{n_\gamma}(z) = {n_\gamma}(0){(1 + z)^3}
\eeq
Then the average separation between photons propagating in the same direction is:
\beq\label{38}
{d_{{\rm{mean}}}}(z) = {n_\gamma }{(z)^{ - 1/3}} \approx 0.31{(1 + z)^{ - 1}}{cm}
\eeq

In Eq. \eqref{26}, ${\Delta z}$ must be negative for photon 2 to suffer from the metric perturbation of photon 1. Moreover, from Eq. \eqref{36} one infers that $\Delta \phi$ is always zero when $\Delta x=0$. This is consistent with the properties of Lorentz transformations. When two photons move in parallel, they do not interact with each other via gravity. 

Considering the above, using ${d_{{\rm{mean}}}}$ as a reference we carefully select nine typical sets of  geodesics with different initial conditions. Their initial values which are adopted in the following numerical simulations can be found in Table \ref{initial}. In our numerical simulations, we adopt Planck 2018 cosmological parameters  \cite{r11}: ${\Omega _\Lambda }=0.6849, {\Omega _m}=0.3150, {\Omega _\gamma }=0.0001$ and  ${H_0}=67.4~\text{km} \cdot \text{s}^{ - 1} \cdot \text{Mpc}^{ - 1}$ to determine the evolution of universe. The scale factor at present, ${a_0}$, is set to 1. Note that we have measured entanglement phase $\Delta \phi$ in units of $ - {w_{1\gamma }}{w_{2\gamma }}$.

\begin{table}[h]\centering
	\begin{threeparttable}
		\caption{Sets of initial values for numerical simulations}
		\begin{tabular}{ccccc}\hline\hline
			Sets &    $\Delta x$                & $\Delta z$                           &  $\Delta \dot x$                \\
			\hline
			S1   &  $1.03 \times {10^{ - 11}}$  &   $-1.03 \times {10^{ - 11}}$        &         0                       \\
			S2   &  $1.03 \times {10^{ - 11}}$  &   $-2.07 \times {10^{ - 11}}$        &         0                       \\
			S3   &  $1.03 \times {10^{ - 11}}$  &   $-3.10 \times {10^{ - 11}}$        &         0                       \\
			S4   &  $1.03 \times {10^{ - 11}}$  &   $-1.03 \times {10^{ - 11}}$        &   $7.17 \times {10^{ - 30}}$    \\
			S5   &  $1.03 \times {10^{ - 11}}$  &   $-1.03 \times {10^{ - 11}}$        &   $1.43 \times {10^{ - 29}}$    \\
			S6   &  $2.07 \times {10^{ - 11}}$  &   $-1.03 \times {10^{ - 11}}$        &         0                       \\
			S7   &  $3.10 \times {10^{ - 11}}$  &   $-1.03 \times {10^{ - 11}}$        &         0                       \\
			S8   &  $1.03 \times {10^{ - 11}}$  &   $-1.03 \times {10^{ - 11}}$        &   $-7.17 \times {10^{ - 30}}$   \\
			S9   &  $1.03 \times {10^{ - 11}}$  &   $-1.03 \times {10^{ - 11}}$        &   $-1.43 \times {10^{ - 29}}$   \\
			\hline\hline
		\end{tabular}\label{initial}
		\begin{tablenotes}
			\item[1] Note that this is in natural unit system. In SI units, $\Delta x$, $\Delta z$ should be multiplied by $c=3\times 10^8 \text{m/s}$. For example, for S1, we have $\Delta x=-\Delta z=3.1 \times 10^{-3} \text{m}$. In order for the perturbation metric to induce entanglement phase, one of photons must lag behind the other by a minus $\Delta z$.
		\end{tablenotes}
	\end{threeparttable}
\end{table}

\begin{figure}
	\centering
	\subcaptionbox{Phase change under different transverse separations $\Delta x$.\label{f4a}}
	{\includegraphics[scale=0.5]{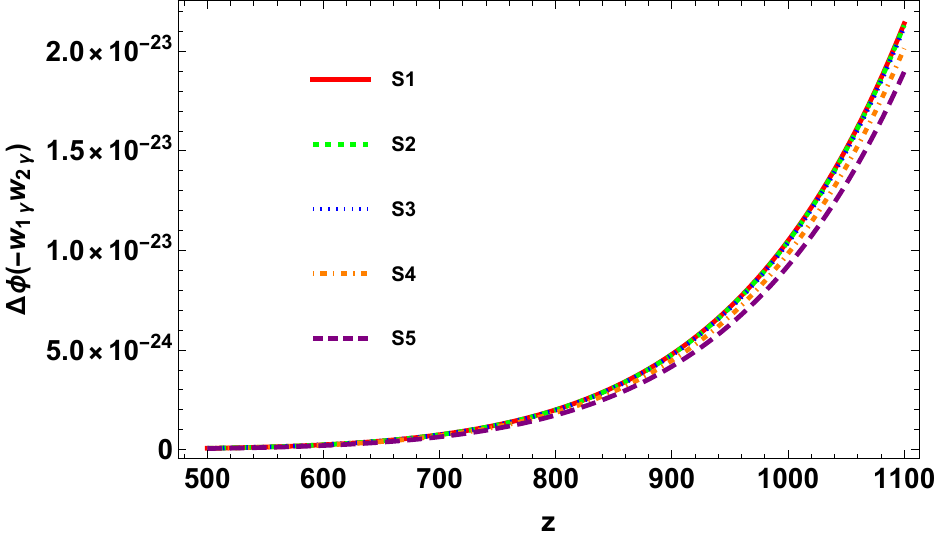}}
	\subcaptionbox{Phase change under different longitudinal separations $\Delta z$.\label{f4b}}
	{\includegraphics[scale=0.5]{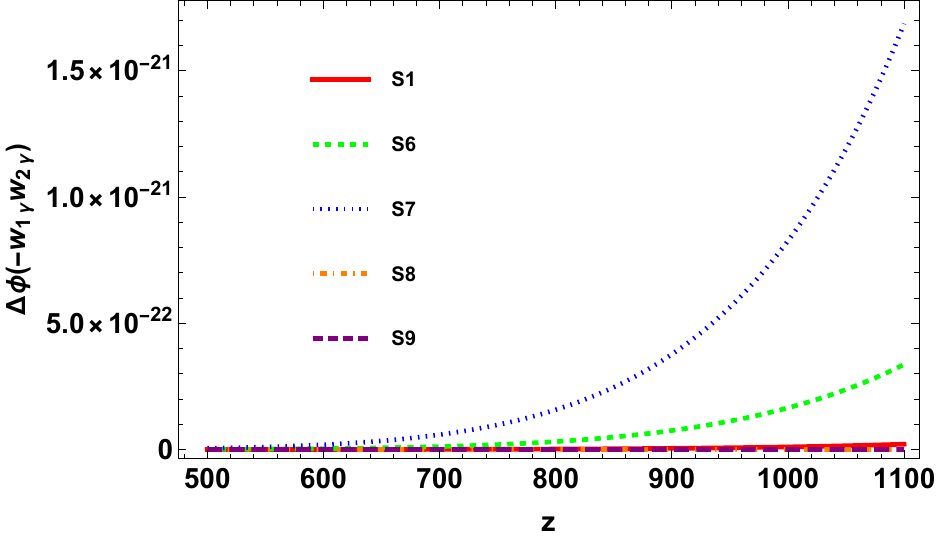}}
	\caption{Phase change ${\Delta {\phi}}$ as a function of z.}\label{f4}
\end{figure}

Fig. \ref{f4} shows that for all initial conditions considered, the entanglement phase of the geodesics increases rapidly with redshift $z$. In Fig. \ref{f4a} the curves for S1, S2 and S3 nearly coincide, indicating a weak dependence on $\Delta z$. The geodesics with a small transverse deviation in the $x$-direction, S4 and S5, exhibit slightly smaller phase changes. This reduction arises from cancellation between positive and negative terms in Eq. \eqref{28}. Nevertheless, the phase is highly sensitive to the $\Delta x$ separation. Because Eq. \eqref{36} gives $\Delta\phi \propto {\Delta x}^{4}$, the geodesics with larger $\Delta x$, S6 and S7, in Fig. \ref{f4b} show substantially larger entanglement phases compared with S1. Similarly, the phase changes for S8 and S9, which have negative deviation velocities, are slightly smaller than that of S1.

\begin{figure}
	\centering
	\includegraphics[scale=0.3]{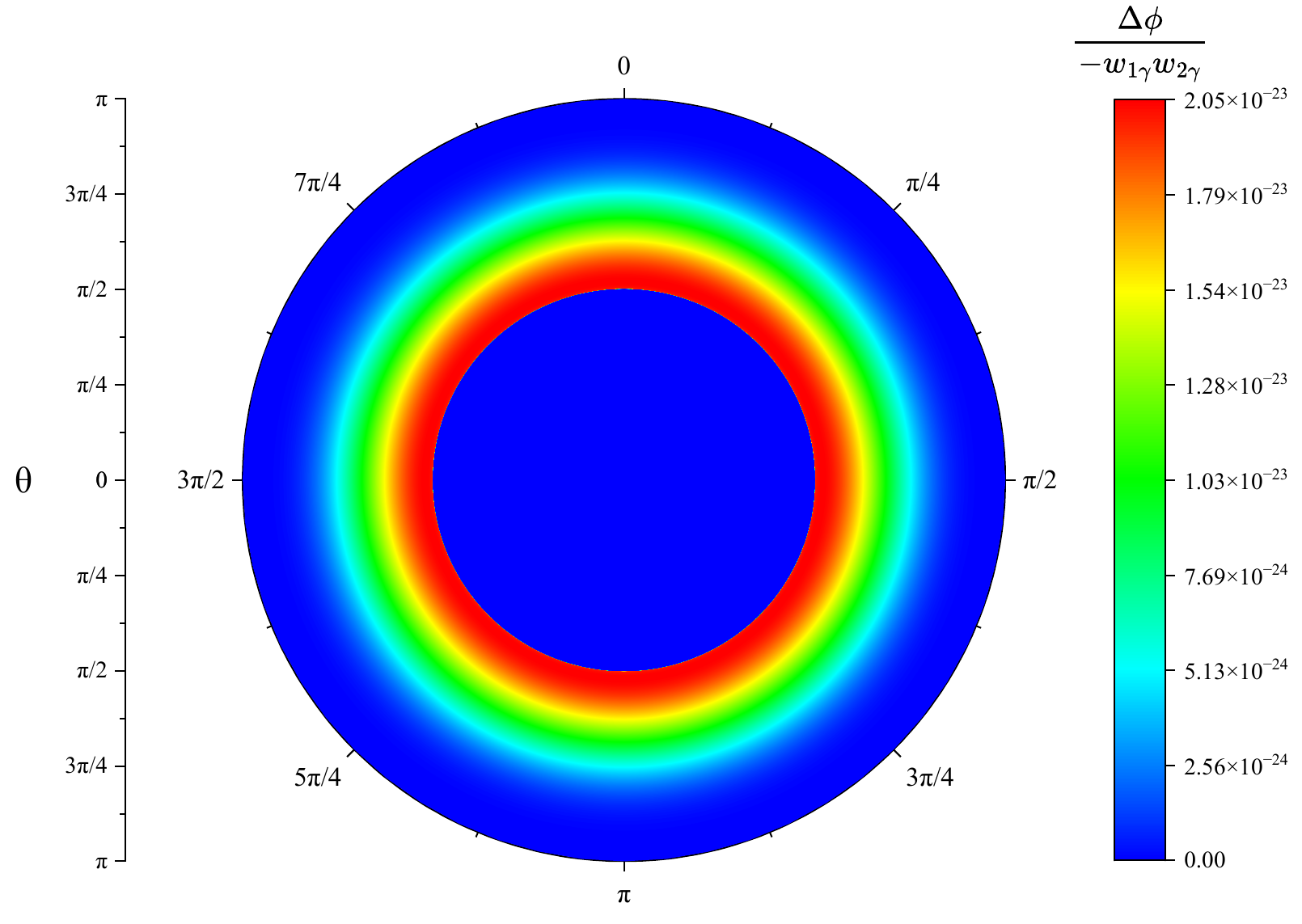}
	\caption{Contour plot of phase change ${\Delta \phi}$ as a function of azimuthal angle $\varphi$ and polar angle $\theta$ with spatial separation fixed at ${d_{{\rm{mean}}}}$. Assume the photon pair is emitted at $z = 1100$. ${\Delta \phi}$ has been measured in units of $ - {w_{1\gamma }}{w_{2\gamma }}$.}\label{f8}
\end{figure}

Unlike Ref. \cite{Zhang:2023bka}, where the entanglement phase depends only on the magnitude of the spatial separation, here the phase acquired by photon pairs exhibit an additional dependence on the direction of the separation vector. We therefore fix the separation distance at ${d_{{\rm{mean}}}}$, assume vanishing geodesic‐deviation velocity to simplify analysis, and compute the entanglement phase as photon 2 locates various angular positions relative to photon 1.  Since CMB photons constitute approximately 99\% of cosmic photons \cite{Dwek:2012nb} and originate from photon decoupling, we henceforth assume the photons of interest were emitted at decoupling corresponding to redshift $z\approx1100$. As illustrated in Fig. \ref{f8}, the entanglement phase $\Delta\phi$ possesses symmetry along the azimuthal angle direction $\varphi$, which is a consequence of the axial symmetry of the metric perturbation. When the polar angle $\theta \le \frac{\pi }{2}$, photon 2 can never enter the light cone of perturbation, and thus $\Delta\phi$ remains zero. Once $\theta$ slightly exceeds $\frac{\pi }{2}$, photon 2 is able to enter the light cone of perturbation, at which point the transverse separation between the two photons reaches its maximum and $\Delta\phi$ attains its peak value. As $\theta$ further increases, the transverse  separation decreases, and $\Delta\phi$ gradually diminishes until it vanishes.

\begin{figure*}\centering
	\subcaptionbox{Random sampling points of $\Delta x$.\label{f10a}}
	{\includegraphics[scale=0.56]{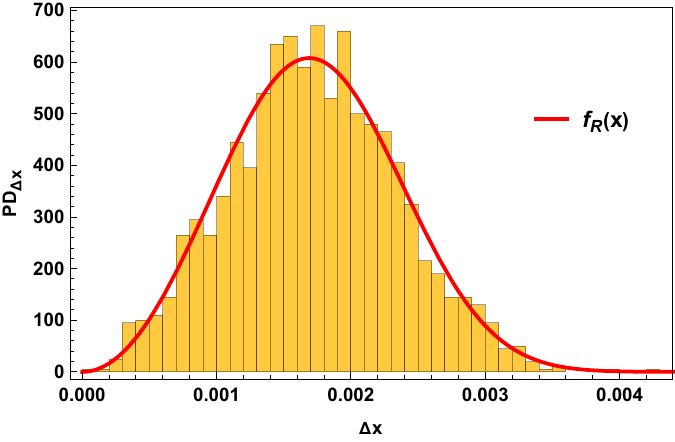}}
	\subcaptionbox{Random sampling points of $\theta $.\label{f10b}}
	{\includegraphics[scale=0.56]{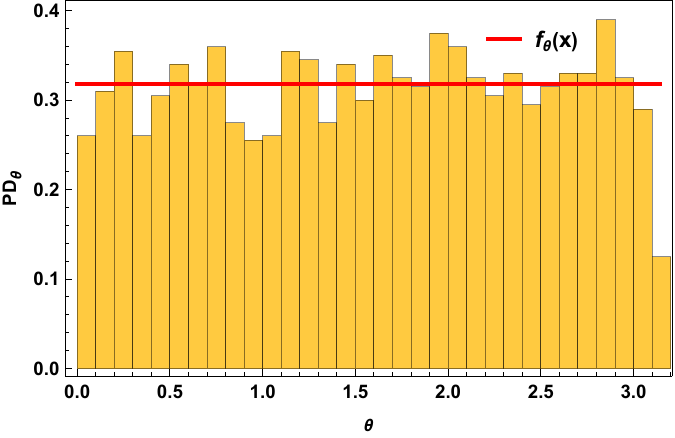}}
	\subcaptionbox{Random sampling points of the momentum of 1 photon, ${p_1}$.\label{f10c}}
	{\includegraphics[scale=0.56]{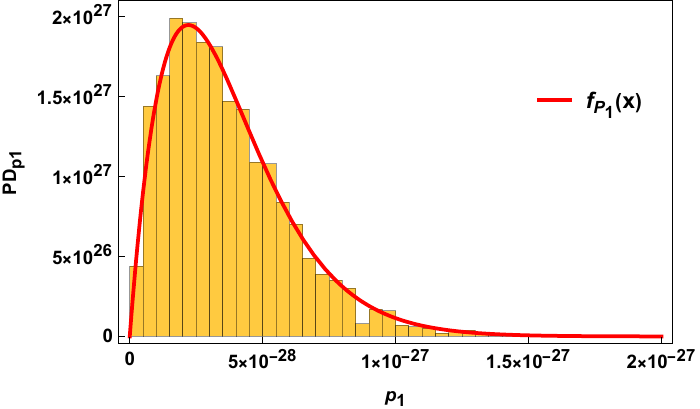}}
	\subcaptionbox{Random sampling points of the momentum of 2 photon, ${p_2}$.\label{f10d}}
	{\includegraphics[scale=0.56]{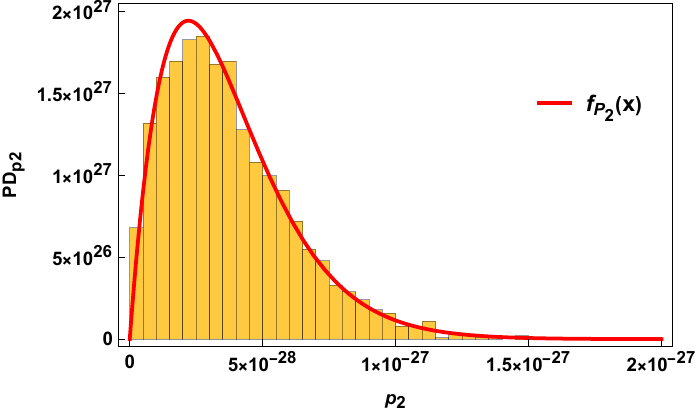}}
	\caption{The distribution of 2000 sample points of the four independent random variables. The horizontal axis denotes the sampled random variables and the vertical axis denotes probability density (PD). The red curve shows the probability density function from which the samples were drawn.}\label{f10}
\end{figure*}

In the above, we investigated the relationship between the gravity-induced entanglement phase of photons and various initial parameters by employing the concept of average spatial separation. In realistic scenarios, however, the distribution of separations between the nearest photons propagating in a given direction follows a Homogeneous Poisson Point Process distribution, with probability density function \cite{Chandrasekhar:1943ws}:
\beq\label{41}
{f_p}\left( {r,\theta ,\varphi } \right) = {r^2}{n_\gamma }\exp \left( { - \frac{{4\pi }}{3}{r^3}{n_\gamma }} \right)\sin \left( \theta  \right)
\eeq
and photon energies are distributed according to the blackbody radiation law:
\beq\label{42}
{f_m}\left( p \right) = \frac{1}{{2\zeta (3)}}{\left( {\frac{1}{{{k_B}T}}} \right)^3}\frac{{{p^2}}}{{{e^{p/({k_B}T)}} - 1}}
\eeq 
So we next employ Monte Carlo methods to simulate the distribution of the entanglement phase in reality. Based on the probability density functions \eqref{41}–\eqref{42} for the independent random variables $\Delta x$, $\theta$, ${p_1}$ and ${p_2}$, we extracted 2000 random sample points, as shown in Fig. \ref{f10}.

\begin{figure}\centering
	\includegraphics[scale=0.5]{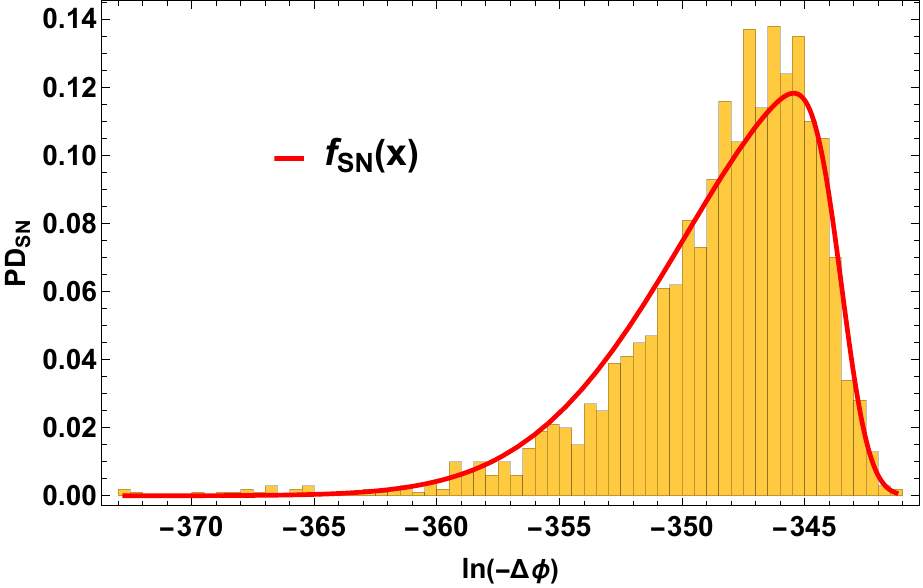}
	\caption{Histogram of Monte Carlo–simulated entanglement phase. Both photons are assumed emitted at $z=1100$. The horizontal axis represents the logarithm of absolute entanglement phase and the vertical axis represents the probability density per bin.}\label{f2}
\end{figure}

Then we analyzed the $\Delta \phi$ values from the 2000 Monte Carlo samples. We now focus solely on the entanglement phase of the delayed photon, without distinguishing between photon 1 and photon 2 anymore. The entanglement phases are extremely small and span many orders of magnitude. So we plot a histogram of their logarithmic values in Fig. \ref{f2}. As model-free statistics, the sample mean and median calculated directly from the raw data are

\begin{widetext}
\beq\label{44}
\begin{split}
&\mathrm{mean}(\Delta \phi)=-1.21\times 10^{-150}\quad\quad\text{95\% CI:} -1.37\times 10^{-150}  \text{ to } -1.07\times 10^{-150}, 
\\
&\mathrm{median}(\Delta \phi)=-1.25\times 10^{-151}\quad\text{95\% CI:} -1.49\times 10^{-151} \text{ to } -1.08\times 10^{-151}.
\end{split}
\eeq
\end{widetext}
The 95\% confidence intervals (CI) above were obtained by a nonparametric bootstrap with \(B=2000\) resamples.

Since the histogram in Fig. \ref{f2} exhibits a skewed, unimodal distribution, we model the logarithm
$x=\ln \left( { - \Delta \phi } \right)$ by a skew-normal distribution \cite{Tarnopolski:2015aja}:
\beq\label{45}
\begin{split}
	&{f_{SN}}(x;\xi ,\omega ,\alpha )= \frac{2}{\omega}\,
	\varphi\!\left(\frac{x-\xi}{\omega}\right)
	\,\Phi\!\left(\alpha\frac{x-\xi}{\omega}\right),
	\\
	&\varphi(z)=\frac{1}{\sqrt{2\pi}}e^{-z^2/2},\quad
	\Phi(z)=\int_{-\infty}^{z}\varphi(t)\,\mathrm{d}t.
\end{split}
\eeq
where $\alpha,\xi,\omega$ denote the shape, location, and scale parameters respectively, and $\varphi$ and $\Phi$ are the standard normal pdf and cdf. The parameter estimates obtained using MLEs (maximum likelihood estimates) in log-space are
\beq\label{46}
\alpha=-6.913,\quad
\xi=-343.5,\quad
\omega=6.328.
\eeq
The 95\% CI by parametric bootstrap are
\beq\label{47}
\begin{split}
&\alpha\in[-8.280,\,-5.941],\quad
\xi\in[-343.7,\,-343.4], \\
&\omega\in[6.091,\,6.558].
\end{split}
\eeq
Physically, modeling $\ln \left( { - \Delta \phi } \right)$ is appropriate here because the data are strictly positive and span many orders of magnitude, a situation typical of multiplicative-generation processes. 

Using the fitted log-space model and a parametric bootstrap with \(B=2000\) replicates, the model-based estimates are:
\begin{widetext}
\beq\label{43}
\begin{split}
&\mathrm{mean}(\Delta \phi)= -1.22\times 10^{-150}\quad\quad \text{95\% CI:} -1.36\times 10^{-150} \text{to}  -1.07\times 10^{-150},
\\
&\mathrm{median}(\Delta \phi)= -8.90 \times 10^{-152}\quad \text{95\% CI:}-1.13\times 10^{-151} \text{to} -6.97\times 10^{-152}.
\end{split}
\eeq
\end{widetext}
These model-based intervals are broadly consistent with the confidence intervals of raw data reported above.

Compared with massive particles \cite{Zhang:2023llk,Zhang:2023bka}, the phase change is extremely tiny for massless photons overall due to the characteristics of the gravitational field of fully relativistic particles. The entanglement phase is so small that there is no practical significance for observation to verify the quantum nature of gravity. Nevertheless, it can  provide useful insights for photon-entanglement experiments conducted in both cosmological and laboratory environments.

\section{Conclusions}

Although the QGEM proposal offers significant value for probing the interaction between gravity and quantum matter, it does not constitute definitive evidence for gravitational field quantization or gravitons, since the original implementation fails to satisfy system locality \cite {Martin-Martinez:2022uio}. In this article, we generalize the QGEM proposal \cite{r12,r13} to relativistic photons in the FLRW universe. It is found that entanglement induced by gravitational interaction between photons propagating in parallel would provide more convincing evidence for the quantum nature of gravitational field, because both LOCC locality conditions are satisfied from the interaction Hamiltonian \eqref{34} and the evolution operator \eqref{35}. 

We then analyzed the influence of the relevant kinematic parameters on the entanglement phase. The result shows that larger redshift promotes the formation of entanglement. The entanglement phase is insensitive to longitudinal separation but exhibits strong sensitivity to transverse separation. When the polar angle between the two photons exceeds $\frac{\pi }{2}$, the entanglement phase decreases with increasing angle until it vanishes. Furthermore, a Monte Carlo simulation of photons produced at the epoch of cosmic decoupling indicates that the logarithmic entanglement phase approximately satisfies a skew-normal probability distribution.

Our study demonstrates that, in contrast to the flat spacetime background, cosmic expansion enables parallel-propagating photons to become entangled via gravitational interaction through the QGEM mechanism. Previous studies  \cite{r4,r5,r6,r7} have typically focused on the entangled states generated through the interaction of photons with other forms of matter, followed by their decoherence during propagation to Earth. In this work, photons are not entangled beforehand but become entangled by gravitational interactions with neighboring photons as they propagate. The results indicate that the QGEM entanglement effects are extremely weak, so we need not worry about obscuring the information relevant to their purpose.

Although the photon entanglement generated via the QGEM mechanism is extremely weak, our study theoretically explores the distinctive features of gravitational interactions that set them apart from other fundamental forces, as manifested through quantum entanglement. Experimentally, our results rule out the possibility of using photons to probe  the quantum nature of first-order radiative gravitational fields beyond Newtonian gravity, while also offering insights for other photon entanglement experiments.

\end{document}